\documentstyle[sprocl]{article}

\def\Journal#1#2#3#4{{#1} {\bf #2}, #3 (#4)}

\def\S{ + \!\!\!\!\!\! \supset }

\def\be{\begin{equation}}
\def\ee{\end{equation}}
\def\bea{\begin{eqnarray}}
\def\eea{\end{eqnarray}}
\begin{document}
\begin{flushright}
{hep-th/0110257}
\end{flushright}

\title{CAYLEY--KLEIN CONTRACTIONS OF ORTHOSYMPLECTIC SUPERALGEBRAS}

\author{N. A. GROMOV, I. V. KOSTYKOV, V. V. KURATOV}

\address{Department of Mathematics, Syktyvkar Branch of IMM UrD RAS, 
Chernova st., 3a, Syktyvkar, 167982, Russia \\E-mail: gromov@dm.komisc.ru}

\maketitle\abstracts{We define a class of orthosymplectic superalgebras
$osp(m;j|2n;\omega)$ which may be obtained from
$osp(m|2n)$ by contractions and analytic continuations in a similar
way as the orthogonal and the symplectic Cayley-Klein algebras
are obtained from the corresponding classical ones. Contractions of
$osp(1|2)$ and $osp(3|2)$ are regarded as an examples. }

Since their discovery \cite{1}\,
\cite{2}\, \cite{3} in 1971 the supersymmetry is used in different 
physical theories such as Kaluza--Klein supergravity \cite{W-86},
supersymmetric field theories of the Wess--Zumino type \cite{K-75},
massless higher-spin field theories \cite{Vas-90}. Recently the secret
theory \cite{B-96} (or  S-theory) that includes superstring theory and its
super p-brane and D-brane \cite{BIK}  generalizations was discussed.
All these theories are build algebraically with the help of some
superalgebra in their base. In this work we wish to present a wide
class of Cayley--Klein (CK) superalgebras which may be used for
constractions of different sypersymmetric models.

\section{\bf $osp(m|2n)$ superalgebra}

Let $e_{IJ} \in M_{m+2n}$ satisfying $(e_{IJ})_{KL}=\delta_{Il}\delta_{JK}$
are elementary matrices. One defines the following
graded matrix
$$
G=\left (
\begin{array}{c|c}
 I_m   &   0   \cr \hline
0  &   0 \quad  I_n   \cr
   &  -I_n \quad 0
\end{array}
\right )
$$
where $I_m,I_n$ are identity matrices. Let $i,j,\ldots=1,\ldots ,m, \,
\bar i,\bar j=m+1, \ldots, m+2n.$
The generators of the orthosymplectic superalgebra $osp(m|2n)$ are given by
$$
E_{ij}=-E_{ji}=\sum_{k}(G_{ik}e_{kj}-G_{jk}e_{ki}),\;\;
E_{\bar i\bar j}=E_{\bar j\bar i}=\sum_{\bar k}
(G_{\bar i\bar k}e_{\bar k\bar j}+
G_{\bar j\bar k}e_{\bar k\bar i}),\;\;
$$
\begin{equation}
E_{i\bar j}=E_{\bar ji}=\sum_{k}G_{ik}e_{k\bar j}+
\sum_{\bar k}G_{\bar j\bar k}e_{\bar ki},
\label{1}
\end{equation}
where the even (bosonic) $E_{ij}$ generate the  $so(m)$ part,
the even (bosonic)
 $E_{\bar i\bar j}$ generate the $sp(2n)$ part and the
 rest $E_{i\bar j}$ are the  odd
 (fermionic) generators of superalgebra. They satisfy
 the following (super) commutation relations
$$
[E_{ij},E_{kl}]=G_{jk}E_{il}+G_{il}E_{jk}-G_{ik}E_{jl}-G_{jl}E_{ik},\;\;
$$
$$
[E_{\bar i\bar j},E_{\bar k\bar l}]=-G_{\bar j\bar k}E_{\bar i\bar l}-
G_{\bar i\bar l}E_{\bar j\bar k}-G_{\bar j\bar l}E_{\bar i\bar k}-
G_{\bar i\bar k}E_{\bar j\bar l},
$$
$$
[E_{ij},E_{k\bar l}]=G_{jk}E_{i\bar l}-
G_{ik}E_{j\bar l},\;\;
[E_{i\bar j},E_{\bar k\bar l}]=-G_{\bar j\bar k}E_{i\bar l}-
G_{\bar j\bar l}E_{i\bar k},
$$
\begin{equation}
[E_{ij},E_{\bar k\bar l}]=0, \quad
\{E_{i\bar j},E_{k\bar l}\}=
G_{ik}E_{\bar j\bar l}-
G_{\bar j\bar l}E_{ik}.
\label{2}
\end{equation}

In the matrix form
$
osp(m|2n)=\{M \in M_{m+2n}|M^{st}G+GM=0\}.
$
If the matrix $M$ has the following form:
$
%\begin{equation}
M=\sum_{i,j}a_{ij}E_{ij} + \sum_{\bar i,\bar j}b_{\bar i\bar j}
E_{\bar i\bar j} + \sum_{i\bar j}\mu_{i\bar j}E_{i\bar j},
%\label{3}
%\end{equation}
$
with $a_{ij}, b_{\bar i \bar j}\in $ {\bf R} or {\bf C} and
$\mu_{i\bar j}$ as the odd nilpotent elements of
Grassmann algebra: $\mu^2_{i\bar j}=0,\,
\mu_{i\bar j}\mu_{i'\bar j'}=-\mu_{i'\bar j'}\mu_{i\bar j},$
then the corresponding supergroup $Osp(m|2n)$ is obtained by the
exponential map $ {\cal M}=\exp M $ and act on (super)vector space
by matrix multiplication ${\cal X}'={\cal M}{\cal X},$ where
${\cal X}^t=(x|\theta)^t,$ $x$
is a $n$--dimentsional even vector and $\theta$ is
a $2m$--dimensional odd vector with odd Grassmann elements.
The form $inv=\sum^n_{i=1}x^2_i+2\sum^m_{k=1}
\theta_{+k}\theta_{-k}=x^2+2\theta^2$
is invariant under this action of orthosymplectic supergroup.

\section{Cayley-Klein orthogonal and symplectic algebras}

Orthogonal $so(m)$ and symplectic $sp(2n)$ algebras are even
subalgebras of $osp(m|2n).$ On the other hand both of these algebras
may be contracted and analytically continued to the set of 
CK orthogonal (and symplectic) algebras.
CK group $SO(m;j)$ is defined as the set of transformations
of vector space ${\bf R}_m(j),$ which preserve the form
$x^2(j)=x_1^2+\sum_{k=2}^{m}[1,k]^2x^2_k, $
where $ [i,k]=\prod^{\max(i,k)-1}_{p=\min(i,k)}j_p,
\, [i,i]=1, $
each parameter $j_k=1,\iota_k,i,$
where $\iota_k$ are nilpotent $ \iota^2_k=0,$ commutative
$\iota_k\iota_p=\iota_p\iota_k \neq 0$
generators of Pimenov algebra ${\bf}P(\iota).$
Cartesian components of vector $x(j)\in {\bf R}_m(j)$
are $x^t(j)=(x_1,j_1x_2, \ldots ,[1,m]x_m)^t, $
as it is easily follows from $x^2(j).$
For $m\times m$ matrix $g(j) \in SO(m;j)$ the transformation
$g(j): {\bf R}_m(j) \rightarrow {\bf R}_m(j)$ means that the vector
$x'(j)=g(j)x(j)$ has exactly the same distribution of
parameters $j$ among its components as $x(j).$
This requirement give an opportunity to obtain the
distribution of parameter $j$ among elements of matrix $g(j),$
i.e. to build the fundamental representation of CK group
$SO(m;j)$ starting from the quadratic form. It is remarkable that the
same distribution of the parameters $j$ is hold for CK Lie
algebra $so(m;j),$ namely
$A_{ik}=[i,k]a_{ik},$ for $A \in so(m;j).$

CK symplectic group $Sp(2n;\omega)$ is defined as the set of
transformations of ${\bf R}_n(\omega) \times {\bf R}_n(\omega),$
which preserve the bilinear form
$S(\omega)=S_1+
\sum_{k=2}^{n}(1,k)^2S_k,$ where $S_k(y,z)=y_kz_{n+k}-y_{n+k}z_k, \,
(i,k)=\prod^{\max(i,k)-1}_{p=\min(i,k)}
\omega_k, \, (i,i)=1, \,
\omega_k=1,\xi_k,i, \, \xi^2_k=0, \,
\xi_k\xi_p=\xi_p\xi_k.$
The distribution of parameters $\omega_k$ among matrix elements
of the fundamental representation
$M(\omega)=\left( \begin{array}{cc}
 H(\omega) & E(\omega) \cr
F(\omega) & -H^t(\omega) \end{array} \right)$
of the CK symplectic algebra $sp(2n;\omega)$ may be obtained as for
orthogonal CK algebras and is as follows:
$B_{ik}=(i,k)b_{ik},$ where $B=H(\omega),E(\omega),F(\omega).$

\section{CK orthosymplectic superalgebras $osp(m;j|2n;\omega)$}

We shall define these superalgebras starting with
the invariant form
\begin{equation}
inv=u^2\sum^m_{k=1}[1,k]^2x^2_k+v^22\sum^{m+n}_{k=m+1}(1,\hat {\bar k}-m)^2
\theta_k\theta_{-k}\equiv u^2x^2(j)+v^22\theta^2(\omega),
\label{4}
\end{equation}
$\hat {\bar k}=\bar k-m, \, \bar k=m+1,\ldots ,m+n; \,
\hat {\bar k}=\bar k-2m, \, \bar k=m+n+1, \ldots ,m+2n,$
which is the natural unificatin of CK orthogonal and
 symplectic forms. The distributions of contraction parameters
$j,\omega $ among matrix elements of the fundamental
representation of $osp(m;j|2n;\omega)$
and transformations of the generators (\ref{1})
 are obtained in a standart CK manner and are as follows:
\begin{equation}
E_{ik}=[i,k]E^*_{ik}, \;\;
E_{\bar i\bar k}=(\hat {\bar i},\hat {\bar k})E_{\bar i\bar k},\;\;
E_{i\bar k}=u[1,i]v(1,\hat {\bar k})E^*_{i\bar k},
\label{5}
\end{equation}
where $E^*$ are generators (\ref{1}) of the starting superalgebra
$osp(m|2n).$ The transformed generators are subject of the
(super) commutation relations:
$$
[E_{ij},E_{kl}] = [i,j][k,l] \left (
{{G_{jk}E_{il}} \over {[i,l]}} + {{G_{il}E_{jk}} \over {[j,k]}} -
{{G_{ik}E_{jl}} \over {[j,l]}} - {{G_{jl}E_{ik}} \over {[i,k]}} \right ),
$$

 $$
[E_{\bar{i}\bar{j}},E_{\bar{k}\bar{l}}] =
-(\hat{\bar{i}},\hat{\bar{j}})(\hat{\bar{k}},\hat{\bar{l}}) \left (
{{G_{\bar{j}\bar{k}}E_{\bar{i}\bar{l}}} \over 
{(\hat{\bar{i}},\hat{\bar{l}})}} +
{{G_{\bar{i}\bar{l}}E_{\bar{j}\bar{k}}} \over 
{(\hat{\bar{j}},\hat{\bar{k}})}} + 
{{G_{\bar{i}\bar{k}}E_{\bar{j}\bar{l}}} \over 
{[\hat{\bar{j}},\hat{\bar{l}})}} +
{{G_{\bar{j}\bar{l}}E_{\bar{i}\bar{k}}} \over 
{[\hat{\bar{i}},\hat{\bar{k}})}} \right ),
$$
$$
[E_{ij},E_{\bar{k}\bar{l}}] = 0, \quad
[E_{ij},E_{k\bar{l}}] =
[i,j][1,k] \left (
{{G_{jk}E_{i\bar{l}}} \over {[1,i]}} -
{{G_{ik}E_{j\bar{l}}} \over {[1,j]}} \right ),
$$
 $$
[E_{i\bar{j}},E_{\bar{k}\bar{l}}] =
-(1,\hat{\bar{j}})(\hat{\bar{k}},\hat{\bar{l}}) \left (
{{G_{\bar{j}\bar{k}}E_{i\bar{l}}} \over {(1,\hat{\bar{l}})}} +
{{G_{\bar{j}\bar{l}}E_{i\bar{k}}} \over {(1,\hat{\bar{k}})}} \right ),
$$
\begin{equation}
\{E_{i\bar{j}},E_{k\bar{l}}\} =u^2v^2
[1,i](1,\hat{\bar{j}})[1,k](1,\hat{\bar{l}}) \left (
{{G_{ik}E_{\bar{j}\bar{l}}} \over {(\hat{\bar{j}},\hat{\bar{l}})}} -
{{G_{\bar{j}\bar{l}}E_{ik}} \over {[i,k]}} \right ).
\label{6}
\end{equation}
For $u=\iota$ or $v=\iota, \iota^2=0$ superalgebra $osp(m|2n)$
is contracted to inhomogeneous superalgebra, which is semidirect sum
$ \{E_{i\bar{j}}\} \S (so(m) \bigoplus sp(2n)),$
with all anticommutators of the odd  generators equal to zero
$\{E_{i\bar{j}},E_{k\bar{p}} \} = 0.$

\section{Examples}
\subsection{Kinematical contractions of $osp(1|2)$}

This contractions was described in detail in \cite{Val-99}
both on the level of (super) commutation relations and as
subrepresentations of the fundamental matrix representations
of $osp(3|2).$ The isomorphism of low dimensional Lie algebras
$sp(2)$ and $so(3)$ are essentially used for contraction
$osp(1|2)$ to $(1+1)$ dimensional Poincare and Galilei
superalgebras. This case is not included in general CK contractions
of the previous section and we give here the fundamental
$3\times 3$ representations of $(1+1)$ Poincare and Galilei
superalgebras which was absent in \cite{Val-99}. For this purpose we
need to introduce the algebra $A_4(\xi),$ which is free generated by
$\xi_1,\xi_2,$ where $\xi_1\xi_2=\xi_2\xi_1, \,
\xi^4_1=\xi^4_2=0.$ If one take the basis
 $X^*_{23}=\displaystyle{\frac{i}{2}}E_{23}, \,
X^*_{12}=\displaystyle{\frac{i}{4}}(E_{33}+E_{22}), \,
X^*_{13}=\displaystyle{\frac{1}{4}}(E_{33}-E_{22}), \,
Q^*_+=E_{12}, \, Q^*_-=E_{13}, $
 and one transform the generators as follows
\begin{equation}
 X_{12}=\omega^2_1X^*_{12}, \, X_{23}=\omega^2_2X^*_{23}, \,
 X_{13}=\omega_1^2\omega_2^2X^*_{13}, \, Q_{\pm}=\omega_1\omega_2Q^*_{\pm},
 \label{7}
 \end{equation}
where $3\times 3$ matrix $E$ are given by (\ref{1}), each
parameter $\omega_k=1,\xi_k,i, \, k=1,2, $
then the (super) commutation relations of $osp(1|2;\omega)$
may be written in the form
$$
[X_{12},X_{13}]=\omega^4_1X_{23}, \,
[X_{13},X_{23}]=\omega^4_2X_{13}, \,
[X_{23},X_{12}]=X_{13},
$$
$$
[X_{12},Q_{\pm}]=\pm\frac{i}{2}\omega^2_1Q_{\mp}, \,
[X_{13},Q_{\pm}]=\frac{1}{2}\omega^2_1\omega_2^2Q_{\mp},
[X_{23},Q_{\pm}]=\pm\frac{i}{2}\omega^2_2Q_{\mp},
$$
$$
\{Q_+,Q_-\}=-2i\omega^2_1X_{12},\,
\{Q_+,Q_+\}=-2(X_{13}+i\omega_2^2X_{12}), \,
$$
\begin{equation}
\{Q_-,Q_-\}=2(X_{13}-i\omega_2^2X_{12}),
\label{8}
\end{equation}
which is coincided with the super commutators (4.51) in \cite{Val-99}
for the standart contractions of $osp(1|2).$
Our designations of generators and contraction parameters
are connected with corresponding of \cite{Val-99} as follows:
$\omega_k=\epsilon_k, X_{12}=K_{21}=H, \, X_{13}=K_{20}=P, \,
X_{23}=K_{01}=K.$ The slight differences in a structure constant
are due to the use of complex $so(3;{\bf C})$ instead
of its anti de Sitter real form $so(2,1).$ The $(1+1)$
super Poincare algebra is obtained for
$\omega_1=\xi_1, \omega_2=1$ (compare with (4.12) in \cite{Val-99}) and
$(1+1)$ super Galilei algebra is given by (\ref{8})
for $\omega_1=\xi_1, \omega_2=\xi_2$
(compare with (4.39) in \cite{Val-99}).
%{\it Remark.} 
The commutators $[A,B]=\xi^kC, \, k=1,2,3, $
are regarded as zero, i.e. $[A,B]=0.$

The Grassmann-hull \cite{Bos-91}  $M(\omega)=2\alpha X_{23}+2\beta X_{12}+
2\gamma X_{13}+\mu Q_++\nu Q_- $ of
$osp(1|2;\omega)$ is represented by the matrix
$$
M(\omega)=\left(
\begin{array}{c|cc}
 0   &   \omega_1\omega_2\mu & \omega_1\omega_2\nu     \cr \hline
-\omega_1\omega_2\nu & -i\omega_2^2\alpha &
-\omega_1^2(i\beta+\omega^2_2\gamma)   \cr
\omega_1\omega_2\mu & \omega_1^2(i\beta-\omega^2_2\gamma)&
i\omega_2^2\alpha
\end{array}
\right),
$$
where $\mu, \nu $ are odd grassmannian elements:
$\mu^2=\nu^2=0, \, \mu \nu=-\nu\mu. $
For this simplest case it is possible to find the
corresponding supergroup $Osp(1|2;\omega)$ explicitly 
\cite{M-93},
namely
$$
{\cal M}(\omega)=\exp M(\omega)=I+\frac{\sinh u}{u}M(\omega)+
\frac{\cosh u-1}{u^2}M^2(\omega)+
$$
$$
+\omega_1^2\omega_2^2\frac{2(1-\cosh u)+u\sinh u}{u^2}\mu\nu A+
\omega_1^2\omega_2^2\frac{u\cosh u-\sinh u}{u^3}\mu\nu B(\omega),
$$
where $u^2=\omega_1^4(\beta^2+\omega_2^4\gamma^2)-\omega_2^4\alpha^2, $
$$
M^2(\omega)=\left(\begin{array}{c|c}
-2\omega_1^2\omega_2^2\mu\nu&
-i\omega_2^2\alpha\mu+\omega_1^2(i\beta-\omega^2_2\gamma)\nu
\cr \hline
i\omega_2^2\alpha\nu-\omega_1^2(i\beta+\omega^2_2\gamma)\mu &
u^2+\omega_1^2\omega_2^2\mu\nu  \cr
i\omega_2^2\alpha\mu-\omega_1^2(i\beta-\omega^2_2\gamma)\nu & 0 
\end{array} \right.
$$
$$
\left.  \begin{array}{c}
i\omega_2^2\alpha\nu-\omega_1^2(i\beta+\omega^2_2\gamma)\mu \cr \hline
0                                          \cr
u^2+\omega_1^2\omega_2^2\mu\nu  \end{array} \right),
$$
$$
A=\left(\begin{array}{c|cc}
0 & 0 & 0 \cr \hline
0 & 1 & 0 \cr
0 & 0 & 1 \end{array} \right), \quad
B(\omega)=\left(\begin{array}{c|cc}
0 & 0 & 0 \cr \hline
0 & -i\omega_2^2\alpha & -\omega_1^2(i\beta+\omega_2^2\gamma) \cr
0 & \omega_1^2(i\beta-\omega_2^2\gamma)&
i\omega_2^2\alpha   \end{array} \right).
$$

In the case of superalgebra $osp(1|4)$ the isomorphism
$sp(4) \cong so(5) $ is used in \cite{Val-99}
and the standart kinematical contractions of $osp(1|4)$
to $(3+1)$ super Poincare and $(3+1)$ super Galilei are
regarded for abstract generators and for embedding
in the fundamental representations of $osp(5|4).$

\subsection{CK contractions of $osp(3|2)$}

This superalgebra has $so(3)$ as even subalgebra therefore
their contractions to the kinematical $(1+1)$
Poincare, Newton and Galilei superalgebras may be
fulfilled according to general CK scheme of the first section.
But unlike of two odd generators of $osp(1|2)$ the superalgebra
$osp(3|2)$ has six odd generators. In the basis 
$X_{ik}=E_{ki}, \, k,i=1,2,3, \, F=\displaystyle{\frac{1}{2}}E_{44}, \,
E=-\displaystyle{\frac{1}{2}}E_{55}, \,
H=-E_{45}, \, Q_k=E_{k4}, \, Q_{-k}=E_{k5}$
the generators are affected by the contraction
coefficients $j_1,j_2$ in the following way
\begin{equation}
X_{ik}\to [i,k]X_{ik}, \quad Q_{\pm k}\to [1,k]Q_{\pm k}
\label{12}
\end{equation}
and $H,F,E $ are remained unchanged.
Then superalgebra $osp(3;j|2)$ is given by
$$
[X_{12},X_{13}]=j_1^2X_{23}, \quad [X_{13},X_{23}]=j_2^2X_{12}, \quad
[X_{23},X_{12}]=X_{13},
$$
$$
[H,E]=2E, \quad [H,F]=-2F, \quad [E,F]=H,
$$
$$
[X_{ik},Q_{\pm i}]=Q_{\pm k}, \quad
[X_{ik},Q_{\pm k}]=-[i,k]^2Q^2_{\pm i}, \;\;  i<k,
$$
$$
[H,Q_{\pm k}]=\mp Q_{\pm k}, \quad
[E,Q_k]=-Q_{-k}, \quad [F,Q_{-k}]=-Q_k,
$$
$$
\{Q_k,Q_k\}=[1,k]^2F, \quad \{Q_{-k},Q_{-k}\}=-[1,k]^2E,
$$
$$
\{Q_k,Q_{-k}\}=-[1,k]^2H, \quad \{Q_{\pm i},Q_{\mp k}\}=\pm [1,k]^2X_{ik}.
$$
The non-minimal Poincare superalgebra is obtained for
$j_1=\iota_1, \, j_2=i $ and has the structure
of the semidirect sum
$T \S (\{X_{23}\}\oplus osp(1|2)),$ where abelian
$T=\{X_{12},X_{13},Q_{\pm 2},Q_{\pm 3}\}$ and
$osp(1|2)=\{H,E,F,Q_{\pm 1}\}.$ The Newton superalgebra
$osp(3;\iota_2|2)=T_2 \S osp(2|2),$ where
$T_2=\{X_{13},X_{23},Q_{\pm 3}\} $ and $osp(2|2) $ is generated by
$X_{12},H,E,F,Q_{\pm 1},Q_{\pm 2}.$
Finally the non-minimal Galilei superalgebra may be presented as
semidirect sums
$osp(3;\iota_1,\iota_2|2)=(T \S\{X_{23}\})\S osp(1|2)=
T \S (\{X_{23}\}\oplus osp(1|2)).$

\section*{Acknowledgments}
NG would like to thank Mariano del Olmo for sending the
copy of paper \cite{Val-99}.
This work was supported by Russian Foundation for Basic
Research under Project 01-01-96433.

%\section*{Appendix}
%We can insert an appendix here and place equations so that they 
%are given numbers such as Eq.~(\ref{eq:app}).
%\be
%x = y.
%\label{eq:app}
%\ee

\end{document}